\documentclass{aa}
\usepackage{graphics}

\begin{document}
\thesaurus{03(11.14.1; 11.09.1, Cen A, NGC 5128; 11.09.4; 13.19.1;
13.19.3; 09.13.2)}

\title{Dense gas in the dust lane of Centaurus~A\thanks{Based on observations collected at the
European Southern Observatory, La Silla, Chile}}

\author{Wolfgang Wild\inst{1}
\and Andreas Eckart\inst{2}}

\offprints{W. Wild (wild@astro.rug.nl)}

\institute{Kapteyn Institute, University of Groningen, Landleven
12, P.O. Box 800, 9700 AV Groningen, the Netherlands
 \and Universit\"at zu K\"oln, I. Physikalisches
Institut, Z\"ulpicherstr. 77, 50937 K\"oln, Germany}

\date{Received 16 February 2000/ Accepted 04 May 2000}

\maketitle

\begin{abstract}
The interstellar medium of Centaurus A (NGC 5128) has been studied
extensively in recent years, using mostly molecular lines tracing
low to medium density gas (500 to several $10^3$ cm$^{-3}$). The
amount and distribution of the dense molecular gas was largely
unknown. Here we present new millimeter data of the HCN(1--0),
CS(2--1), and CS(3--2) rotational transitions towards the nearby
radio galaxy Centaurus~A observed with the SEST on La Silla,
Chile. We obtained spectra of the HCN(1--0) emission which traces
dense 10$^4$ cm$^{-3}$ molecular gas at the center and along the
prominent dust lane at offset positions $\pm$60$^{\prime \prime}$
and $\pm$100$^{\prime \prime}$. We also obtained a few spectra of
CS(2--1) and (3--2) tracing densities of $\sim$10$^5$ cm$^{-3}$.
The emission in these lines is weak and reaches a few mK at the
available angular resolutions of
 54$^{\prime \prime}$ - 36$^{\prime \prime}$.
At the central position, the integrated intensity ratio
I(HCN)/I(CO) peaks at 0.064, and decreases to $\sim$0.02 to 0.04
in the dust lane.

Using the new high density tracer data, we estimate the amount,
distribution and physical conditions of the dense molecular gas in
the dust lane of Centaurus A. We find that Cen~A and the Milky Way
are comparable in their HCN(1-0) line luminosity. However, towards
the nucleus the fraction of dense molecular gas measured via the
line luminosity ratio L(HCN)/L(CO) as well as the star formation
efficiency $L_{\rm FIR}/L_{\rm CO}$ is comparable to
ultra-luminous infrared galaxies (ULIRGs). Within the off-nuclear
dust lane and for Cen~A as a whole these quantities are between
those of ULIRGs and normal and infrared luminous galaxies. This
suggests that most of the FIR luminosity of Centaurus~A originates
in regions of very dense molecular gas and high star formation
efficiency.
\end{abstract}

\section{Introduction}
Centaurus~A (NGC 5128) is the closest radio galaxy (distance 3.5
Mpc, 1$^{\prime \prime}=$ 17 pc, Soria et al. 1996, Hui et al.
1993, Israel 1998, Ebneter \& Balick 1983, de Vaucouleurs 1979)
and exhibits a very prominent dust lane. Centaurus A is a strong
radio galaxy with a milliarcsecond nuclear continuum source
(Kellermann et al. 1997; Shaffer \& Schilizzi 1975; Kellermann
1974) and two giant radio lobes. Absorption against the nuclear
source has been found in H{\sc I} (van der Hulst et al. 1983)
and many molecular species and transitions (Gardner
\& Whiteoak 1976; Whiteoak \& Gardner 1971; Bell \& Seaquist 1988;
Seaquist \& Bell 1986, 1990; Phillips et al. 1987; Eckart et al.
1990a; Israel et al. 1990, 1991; Wild et al. 1997;
Wiklind \& Combes 1997).

Several studies of Centaurus~A in the millimeter wavelength range
have been carried out. Although other elliptical galaxies with
dust lanes have recently been detected in CO (Sage \& Galleta
1993), Centaurus is the best object for a detailed study due to
its proximity and corresponding large angular size. Phillips et
al. (1987) and Quillen et al. (1992) observed several positions in
the CO(2-1) line along the dust lane at a resolution of
30$^{\prime \prime}$. In previous papers we presented a fully
sampled map of the $^{12}$CO(1-0) emission together with {\it
IRAS} observations of the FIR continuum (Eckart et al. 1990b),
measurements of the millimeter absorptions lines towards the
nucleus (Eckart et al. 1990a; see also Wiklind \& Combes 1997),
and a $^{12}$CO(2-1) map along the dust lane at a resolution of
22$^{\prime \prime}$ (Rydbeck et al. 1993).

In  Wild et al. (1997) we presented the first fully
sampled $^{13}$CO(1-0) map along the dust lane of Centaurus~A, as
well as single spectra of the $^{13}$CO(2-1) emission in the disk
and C$^{18}$O(1-0) emission at the central position. These new
data allowed us, in combination with the $^{12}$CO(1-0) and
$^{12}$CO(2-1) maps obtained earlier, to study the excitation
conditions of the molecular gas in detail throughout the dust
lane. Using different CO line ratios and their variation across
the disk of Centaurus A, we inferred the physical parameters of
the molecular ISM and their  spatial variations.

Here we extend our investigation to the very dense
(10$^4$ to 10$^5$cm$^{-3}$)
phase of the molecular gas by observing
line emission of the density tracers HCN and CS
towards the nucleus and the off-nuclear dust lane in Centaurus~A.

\begin{table}[b]
\caption{Observational parameters} \label{obstab}
\begin{tabular}{lcrrr}\hline \hline
Transition & offset  & integration & $T^{*}_{\rm A,max}$ & $\Delta
v$
\\

     & RA, Dec  & time \\

     & (arcsec) & (min) & (mK) & (km/s) \\

\hline

HCN(1-0) & +105,-73 & 178 & $<$2.0 &  \\

 & +79,-55  & 460 & 4.0 & 230 \\

 & +50,-25  & 396 & 5.0& 190 \\

 & 0,0  & 252 & 8.5 & 390 \\

 & -52,+37 & 836 & 3.8 & 170 \\

 & -79,+55  & 360 & 4.0 & 170 \\

 \hline

 CS(3-2) & +50,-25  & 1644 & 0.9 & 250 \\

   & 0,0 & 712 & 3.8 & 105 \\

CS(2-1) & -52,+37 & 288 & 3.0 & 135 \\

\hline \hline
\end{tabular}
\end{table}
\begin{figure}
\resizebox{\hsize}{!}{\includegraphics{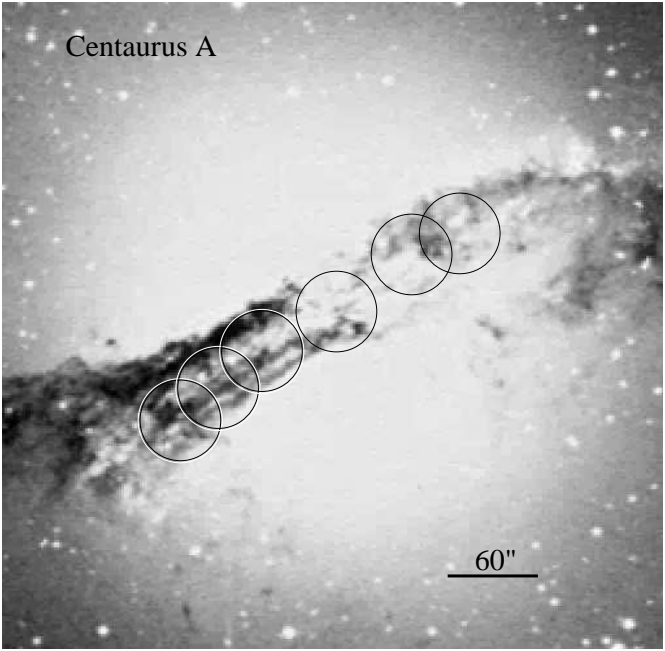}} \hfill
\caption{ Beam positions at which the dense molecular gas in the
dust lane of Centaurus~A has been investigated. Shown is the beam
for the HCN J=1--0 measurements.} \label{figbeams}
\end{figure}
\begin{figure}
\resizebox{\hsize}{!}{\includegraphics{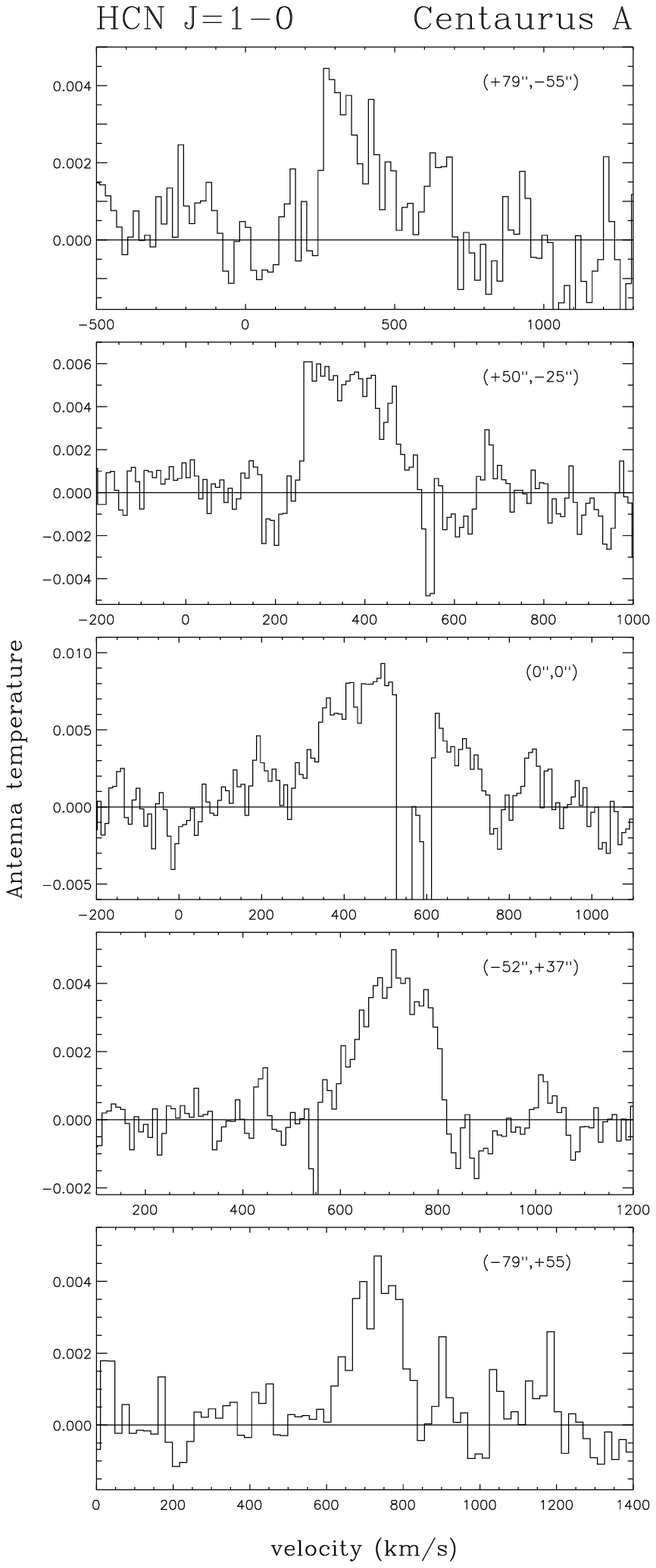}} \hfill
\caption{ HCN J=1--0 spectra along the dust lane of Centaurus~A.
Offsets are given relative to the center position
$\alpha(1950)=13^{h}22^{m}31.8^{s}$,
$\delta(1950)=-42^{\circ}45^{\prime}30^{\prime \prime}$ }
\label{fighcn}
\end{figure}
\begin{figure}
\resizebox{\hsize}{!}{\includegraphics{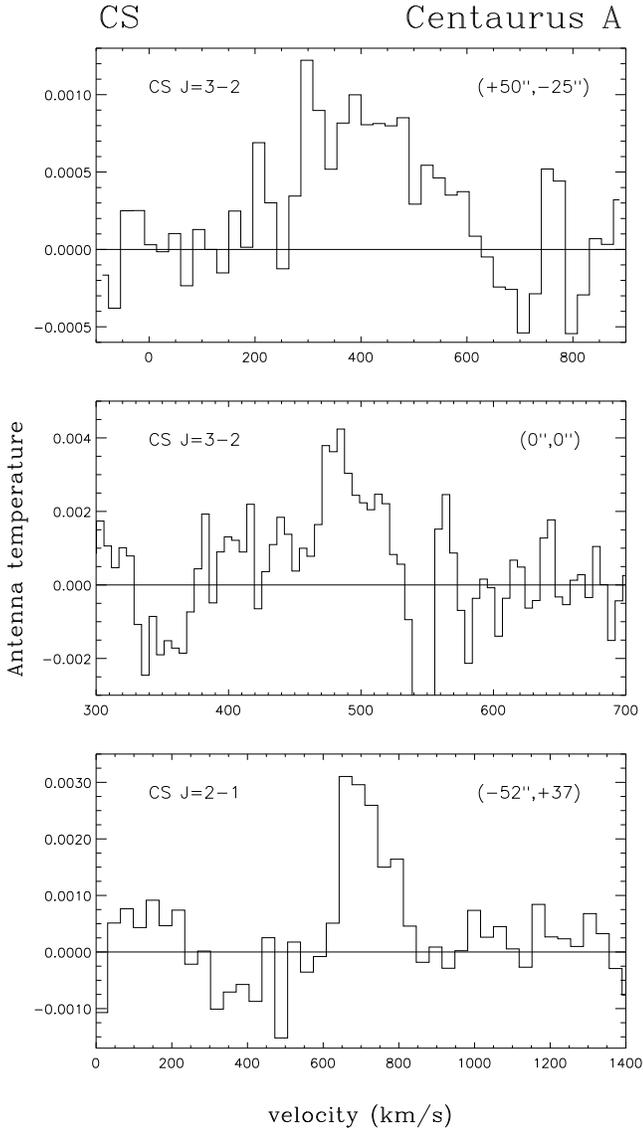}} \hfill
\caption{ CS J=3--2 and J=2--1 spectra along the dust lane of
Centaurus~A. Offsets are given relative to the center position
$\alpha(1950)=13^{h}22^{m}31.8^{s}$,
$\delta(1950)=-42^{\circ}45^{\prime}30^{\prime \prime}$ }
\label{figcs}
\end{figure}

\section{Observations}

The observations were carried out in two observing runs in January 1996
and January 1998 with the 15m Swedish
 ESO Submillimeter
Telescope (SEST) on La Silla, Chile. The adopted central position
for Centaurus A was $\alpha(1950)$ $= 13^{h}22^{m}31.8^{s}$ and
$\delta(1950) = -42^{\circ}45^{\prime}30^{\prime \prime}$. We
observed the $^{12}$CO(1-0), HCN(1-0), CS(2-1), and CS(3-2)
molecular line emission and absorption towards the central
non-thermal continuum source and selected positions in the dust
lane. We used the 3~mm and 2~mm SIS receivers with system
temperatures around 140~K and 180~K, respectively, and beam widths
ranging from  34$^{\prime \prime}$ (FWHM) at 147~GHz to
56$^{\prime \prime}$ at 89~GHz. We observed in the dual beam
switching mode, i.e. a chopper wheel switched between two
positions on the sky displaced by about 12 arcminutes in azimuth.
First the source was placed in one beam and then in the other
beam. The backend was the low resolution acousto-optical
spectrometer with a bandwidth of about 1 GHz in 1440 channels. The
separation between channels was 0.7~MHz, and the actual spectral
resolution 1.4~MHz (4.7 km/s at 89 GHz). We adopted the LSR
velocity scale. The system was flux calibrated with the chopper
wheel method (Kutner \& Ulich 1981). We used main beam
efficiencies $\eta_{\rm \, MB, 89 GHz}$ = 0.75, $\eta_{\rm \, MB,
97 GHz}$ = 0.73, $\eta_{\rm \, MB, 115 GHz}$ = 0.70, and
$\eta_{\rm \, MB, 147 GHz}$ = 0.66 for the conversion from antenna
temperature to Rayleigh-Jeans main beam brightness temperatures.
The pointing was accurate to within 3$^{\prime \prime}$. It was
checked frequently on the nuclear continuum source of Centaurus~A,
and the nearby (distance $\sim 17^{\circ}$) SiO maser W Hya.
Fig.~\ref{figbeams} shows the positions in the dust lane of
Centaurus~A at which the dense molecular gas has been
investigated. Fig.~\ref{fighcn} and Fig.~\ref{figcs} show the HCN
J=1--0 and CS spectra along the dust lane. Table~\ref{obstab}
gives parameters of the observed spectra.

\section{Analysis}

In order to measure the mass of dense molecular gas and therefore
the star formation potential in the dust lane of Centaurus~A we
observed HCN and CS line emission. Both molecules trace higher
density gas than CO, since they have large dipole moments and
require n(H$_2$)$\ge$10$^4$cm$^{-3}$ for significant excitation.
The analysis is done by calculating the line luminosities and
estimating from them the amount of dense molecular gas. These
properties are then discussed and compared to data from the Milky
Way and external galaxies.

Line luminosities can be obtained from the integrated line
intensities $I_{\rm \, line}=\int T_{\rm MB} \times \delta v$ via

\begin{equation}
L_{\rm \, line} = I_{\rm \, line}~D^2~\Omega~~~,
\end{equation}

where $T_{\rm MB}$ is the main beam brightness temperature, D is
the distance to the source and $\Omega$ is the solid angle of the
beam convolved with the source.

\begin{table*}[htb]
\caption{HCN/CO Line Ratios in Centaurus~A}
\label{ratiotab}
\begin{center}
\begin{tabular}{crrrrr}\hline \hline
          &          &          &          &            \\
offset    & distance & I(CO)    & I(HCN) &$\frac{\rm I(HCN)}{\rm
I(CO)}$
\\ (RA, Dec) & from center & (K~km/s) & (K~km/s) &            \\
(arcsec)  & (arcsec) &          &          &            \\ \hline
          &          &          &          &            \\
+105, -73 &  128     &     16.4 &  $<$0.40 &$<$0.024    \\
 +79, -55 &   96     &     34.4 &     1.27 &   0.037    \\
 +50, -25 &   56     &     63.1 &     1.27 &   0.020    \\
   0,~0   &    0     &     69.0 &     4.40 &   0.064    \\
 -52, +37 &  -64     &     50.1 &     0.85 &   0.017    \\
 -79, +55 &  -96     &     27.3 &     0.91 &   0.033    \\
          &          &          &          &            \\
\hline
\hline
\end{tabular}
\end{center}
Intensities are given in $\int T_{\rm MB} \times \delta v$.
\\
Last column gives ratios. The assumed main beam efficiencies are
$\eta_{\rm \, MB, 89 GHz}$=0.75 and $\eta_{\rm \, MB, 115
GHz}$=0.70.
\end{table*}
\begin{table*}[htb]
\caption{CO, HCN, and FIR luminosities} \label{lumtab}
\begin{center}
\begin{tabular}{rrrrrrrrr}\hline \hline
          & &       &     &      &       \\
 distance & L$_{\rm CO}$ & L$_{\rm HCN}$ & L$_{\rm FIR}$ &
 $\frac{\rm L_{FIR}}{\rm L_{CO}}$ &$\frac{\rm L_{HCN}}{\rm L_{CO}}$ \\
 from center    & 10$^6$(L$_{\rm I}$) & 10$^6$(L$_{\rm I}$)& 10$^9$ L$_{\odot}$ &  &   \\
 (arcsec) &          &           &     &  & \\ \hline
          &       &        &     &       &     &      &       \\
  128     &  7.5&  $<$0.31&  0.6&   80 & $<$0.04\\
   96     & 15.7& 1.00 &  0.6&   38 &    0.06\\
   56     & 29.0& 1.00 &  1.2&   41 &    0.03\\
    0     & 31.7& 3.47 &  1.8&   57 &    0.11\\
  -64     & 23.0& 0.67 &  1.2&   52 &    0.03\\
  -96     & 12.5& 0.72 &  0.6&   48 &    0.06\\ \hline
          &     &       &     &      &       \\
total     &  92&    52&  6.0&   65 &    0.06 \\
          & &       &     &      &       \\
\hline \hline
\end{tabular}
\end{center}
$L_{\rm I} = {\rm (K~km~s^{-1})^{-1}~pc^2}$
\\
Intensities are given in $\int T_{\rm MB} \times \delta v$. The
assumed main beam efficiencies are $\eta_{\rm \, MB, 89 GHz}$=0.75
and $\eta_{\rm \, MB, 115 GHz}$=0.70.
\end{table*}

Using radiative transfer solutions (e.g. Kwan \& Scoville 1975,
Linke \& Goldsmith 1980), assuming that the HCN and CS line
emission traces gravitationally bound or virialized clouds with a
density range of a few 10$^4$ cm$^{-3}$ to a few 10$^5$ cm$^{-3}$
and kinetic temperatures between 10 to 60~K. Solomon et al. (1990,
1992) present relations between the mass of molecular gas at these
densities and the corresponding observed line luminosity as
defined above. For HCN(1-0) they find in their 1992 paper
\begin{equation}
M_{\rm HCN} \sim 20^{+30}_{-10} \times L_{\rm HCN}~~~~~~{\rm
M_{\odot}~(K~km~s^{-1})^{-1}}
\end{equation}
and for CS(2-1) they derive in the 1990 paper
\begin{equation}
M_{\rm CS} \sim (35-150) \times L_{\rm CS}~~~~~~{\rm
M_{\odot}~(K~km~s^{-1})^{-1}}~~~.
\end{equation}

The assumptions made above are likely to be valid for the
molecular ISM in Centaurus~A as well. This can be based on the
fact that, according to radiative transport calculations (Eckart
et al. 1990a), the lower density molecular gas - as traced by CO
isotopic emission - has similar properties compared to those in
Galactic GMCs and galaxies with indications for enhanced star
formation activity.

Although the mass estimates derived using the relations given above
include considerable uncertainties
they can be used as a general guide line to investigate the
properties of the dense interstellar medium. The assumed kinetic
temperature range appears to be appropriate for the dense star
forming ISM in Centaurus~A. Eckart et al. (1990b), Joy et al.
(1988), as well as Marston \& Dickens (1988) give dust
temperatures in the dust lane in the range of 40~K. From the
HCN(1-0)/HNC(1-0) ratio of about 2, Israel (1992) introduces
kinetic temperatures of the order of $\sim$25~K.

Table~\ref{ratiotab} lists the integrated line intensities $\int
T_{\rm MB} \times \delta v$ for the HCN(1-0) and the CO(1-0) line
emission, as well as the ratios I(HCN)/I(CO) corrected for main
beam efficiencies. In Table~\ref{lumtab} we list the CO(1-0) and
HCN(1-0) line luminosities as well as the estimated FIR
luminosities $L_{\rm FIR}$ as a function of position and as
estimated quantities integrated over the dust lane.
Table~\ref{CStab} lists line intensities $\int T_{\rm MB} \times
\delta v$ and luminosities (corrected for main beam efficiency)
for CS(2-1) and CS(3-2) lines we measured at a few positions.
Table~\ref{masstab} lists the mass of dense molecular gas as
derived from the line luminosities and Eqs. (2) and (3).

\begin{figure}
\resizebox{\hsize}{!}{\includegraphics{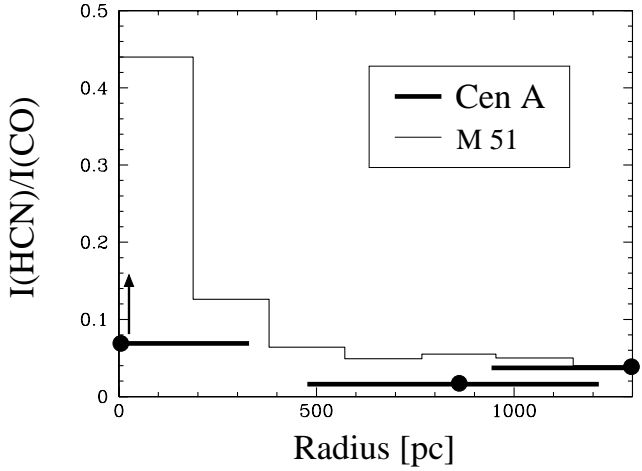}} \hfill
\caption{ Comparison of the HCN(1-0) to CO(1-0) line intensity
ratio in M51 (Kohno et al. 1996) and Cen~A as a function of
position. The central ratio in Cen~A is of the order of 0.1.
Assuming than the HCN(1-0) line emission is concentrated on the
nucleus the HCN(1-0) to CO(1-0) line intensity ratio  will
probably rise (as indicated by the arrow) if high angular
resolution interferometric observations become possible. }
\label{cenam51}
\end{figure}
\begin{figure}
\resizebox{\hsize}{!}{\includegraphics{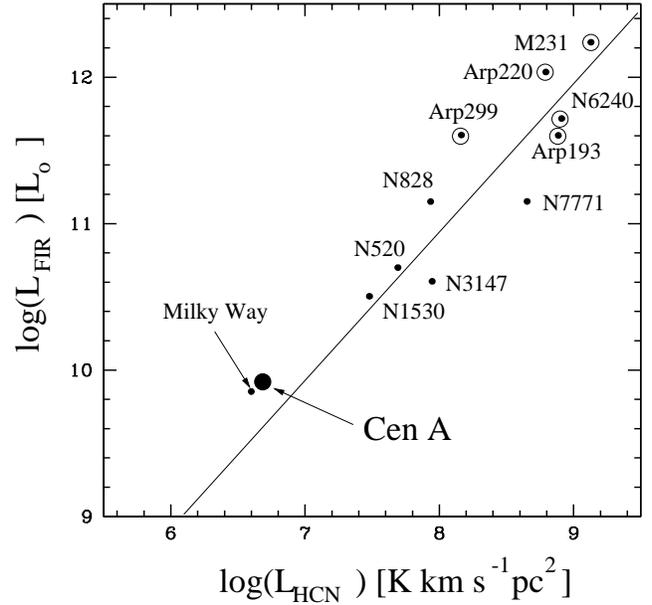}} \hfill
\caption{ Centaurus~A has the same ratio of FIR to HCN luminosity
as ULIRGs (dots in circles) and normal spirals (simple dots;
Solomon et al. 1992). In both quantities Cen~A and the Milky Way
are very similar. } \label{firhcn}
\end{figure}
\begin{figure}
\resizebox{\hsize}{!}{\includegraphics{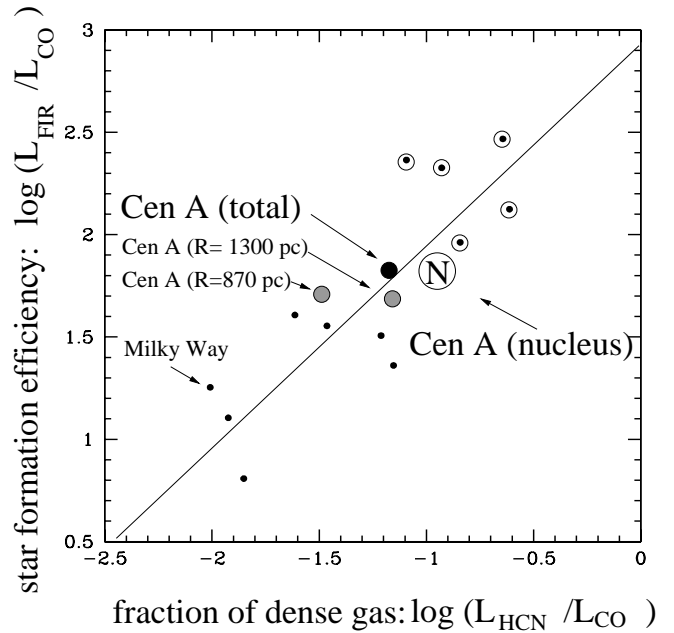}} \hfill
\caption{ Comparison of the fraction of dense gas to the star
formation efficiency (Solomon et al. 1992). The nucleus of Cen~A
has properties of ULIRGS and the bulk of the dust lane falls
between ULIRGS and normal or interacting galaxies. Symbols are the
same as in Fig.~\ref{firhcn}.} \label{fraction}
\end{figure}

\begin{table}[t]
\caption{CS line intensities and luminosities} \label{CStab}
\begin{center}
\begin{tabular}{rrrrrrrrr}\hline \hline
          &               &             &                   &                   \\
 distance & $I_{\rm CS(2-1)}$ &$I_{\rm CS(3-2)}$&$L_{\rm CS(2-1)}$  & $L_{\rm CS(3-2)}$     \\
 from ctr & K~km~s$^{-1}$ &K~km~s$^{-1}$&10$^5$$L_{\rm I}$    & 10$^5$$L_{\rm I}$             \\
 (arcsec) &               &             &                   &                   \\ \hline
   56     &               & 0.33        &                   & 0.9$\times$10$^5$ \\
    0     &               & 0.61        &                   & 1.7$\times$10$^5$ \\
  -64     & 0.55          &             & 3.5$\times$10$^5$ &                   \\ \hline
          &               &             &                   &                   \\
total     &               &             &       --          &
5.4$\times$10$^5$\\
          &               &             &                   &                   \\
\hline \hline
\end{tabular}
\end{center}
Intensities are given in $\int T_{\rm MB} \times \delta v$
corrected for $\eta_{\rm \, MB, 147 GHz}$=0.66 and $\eta_{\rm \,
MB, 97 GHz}$=0.73.
\\
Line luminosities are given in $L_{\rm I} = {\rm
(K~km~s^{-1})^{-1}~pc^2}$.
\\
The total CS(3-2) line luminosity has been derived assuming that
the off-center measurements
 are representative for all of the
dust lane which is 180$^{\prime \prime}$ in length in the CO(1-0)
map (Eckart et al. 1990a).
\end{table}

\section{Results and Discussion}

\subsection{Dense gas traced by HCN}

HCN traces molecular gas at much higher density of about
$\sim$10$^4$cm$^{-3}$ than CO ($\sim$500cm$^{-3}$). In Cen~A the
HCN emission within the dust lane is clearly detected over a
similar velocity range as CO ($v_{\rm LSR}$=250 - 850 km/s). The
HCN emission, however, is stronger peaked on the nucleus compared
to the CO emission. In Fig.~\ref{cenam51} we compare the spatial
variation of the intensity ratio
 I(HCN)/ I(CO) as it is observed in Cen~A and M~51 (Kohno et al. 1996).
At the central position, the integrated intensity ratio I(HCN)/
I(CO) peaks at 0.06, and decreases to $\sim$0.02 to 0.04 in the
dust lane. Ratios of the order of 0.1 or higher are only observed
in active nuclear regions of Seyfert galaxies and ULIRGs (Kohno et
al. 1996; Solomon et al. 1992; Nguyen-Q-Rieu et al.
1992; Helfer \& Blitz 1993, 1995; Jackson et al. 1993; Tacconi et
al. 1994; Sternberg et al. 1994).

The ratio of HCN to CO luminosity is 1/6 for ULIRGs, but only 1/80
in normal spirals (Solomon et al. 1992). For Cen~A the HCN to CO
luminosity is $L_\mathrm{CO}/ L_\mathrm{HCN}$=1/13 and therefore
closer to the value for ULIRGs than  normal spirals.

\begin{table}[htb]
\caption{Dense molecular gas mass traced by HCN and CS}
\label{masstab}
\begin{center}
\begin{tabular}{rrrrrrrrr}\hline \hline
          &    &     \\
 distance from& $M_{\rm HCN}(H_2)$   & $M_{\rm CS}(H_2)$ \\
 center (arcsec)& 10$^7$M$_{\odot}$      & 10$^7$M$_{\odot}$   \\
  &    &     \\
  \hline
          &    &     \\
  128     & $<$0.6&     \\
   96     & 2.0&     \\
   56     & 2.0&  0.8\\
    0     & 4.3&  1.6\\
  -64     & 1.3&     \\
  -96     & 1.4&     \\ \hline
          &    &     \\
total     & 8.6 &  5.0\\
          &    &     \\
\hline \hline
\end{tabular}
\end{center}
Masses have been derived via line luminosities and Eqs. 2 and 3.
Systematic errors are +150\% and -50\% for $M_{\rm HCN}(H_2)$ and
$\pm$40\% for $M_{\rm CS}(H_2)$. The total masses have been
derived assuming that the off-center measurements are
representative for all of the dust lane which is 180$^{\prime
\prime}$ in length in the CO(1-0) map (Eckart et al. 1990a).
\end{table}

Fig.~\ref{firhcn} shows that Centaurus~A has the same ratio of FIR
to HCN luminosity as ULIRGs and normal spirals, including the
Milky Way. This is also true for the center and the off-positions
in the dust lane. This suggests that in Cen~A the star formation
rate per mass of {\it dense} gas is in good approximation
independent on the position and infrared luminosity within the
dust lane.

Detailed observations (Lee et al. 1990) of Galactic GMC's show
that the average CS(2-1)/CO(1-0) intensity ratio is $\sim$1/300.
Since HCN(1-0) is usually 1.5 to 2 times stronger than CS(2-1) and
has larger source sizes Solomon et al. (1992) adopt an
HCN(1-0)/CO(1-0) ratio of $\sim$1/100 for the molecular ring in
the Galactic disk. This results in a HCN luminosity for the Milky
Way of about 4$\times$10$^6$ K~km~s$^{-1}~$pc$^2$ which is quite
comparable to what we find for Centaurus~A with  5.5$\times$10$^6$
K~km~s$^{-1}~$pc$^2$.

In Fig.~\ref{fraction} we show the position of Centaurus~A in a
plot of $log(L_{\rm HCN}/L_{\rm CO})$ which measures the fraction
of dense gas and $log(L_{\rm FIR}/L_{\rm CO})$ which measures the
efficiency with which molecular gas is transformed into OB stars
(Solomon et al. 1992). This plot includes ULIRGs as well as normal
and infrared luminous galaxies. We find that as a whole Cen~A fits
very well into this correlation and is located between both
groups. This is also true for the disk at angular separations of
about 60$^{\prime \prime}$ and 90$^{\prime \prime}$ from the
center.
 However, towards the nucleus the fraction
of dense molecular gas measured via the line luminosity ratio
L(HCN)/L(CO) as well as the star formation efficiency L$_{\rm
FIR}$/L$_{\rm CO}$ is more comparable to ULIRGs rather than normal
and infrared luminous galaxies. This suggests that most of the FIR
luminosity of Centaurus~A originates in regions of very dense
molecular gas and high star formation efficiency.

The HCN line luminosity can now be used to estimate the amount of
dense (10$^4$ cm$^{-3}$) gas via Eq. (2). From the HCN line
luminosity we find a total mass of molecular gas that must be at
densities $>$10$^4$cm$^{-3}$ of $8.6^{+13}_{-4} \times
10^7$M$_{\odot}$. This has to be compared to the total molecular
gas mass derived from the $^{12}$CO(1-0) line of
2$\times$10$^8$M$_{\odot}$ (Eckart et al. 1990a). The comparison
shows that a large fraction - approximately one third - but at
least one sixth -
 of the molecular line emission in Cen~A must originate from
sites with abundant dense molecular gas. The ratio of molecular
gas mass at densities of $>$10$^4$cm$^{-3}$ to gas at
$\sim$300cm$^{-3}$ is about 1/20 in the Galaxy and almost unity
for ULIRGs (Solomon et al. 1992). Based on this quantity the
physical conditions of the dense molecular ISM in Cen~A are
apparently closer to those of ULIRGs than to our Milky Way. The
presence of a large amount of dense molecular gas is also
supported by the fact that if one multiplies $M_{\rm HCN}(H_2)$ by
the mean Galactic $L_{\rm FIR}/M_{\rm HCN}(H_2)$ ratio of
71~$M_{\odot} {\rm (K~km~s^{-1})^{-1}~pc^2}$ (Solomon et al. 1992)
one obtains about 6$\times$10$^9$L$_{\odot}$\ which equals the
observed FIR luminosity. The implicite assumption here is that the
mean Galactic conversion factor is applicable. Accepting this, the
result implies that active star formation in the dust lane of
Cen~A is the actual source of its FIR luminosity and that the AGN
which is responsible for strong radio and X-ray radiation is not
contributing substantially.

\subsection{Dense gas traced by CS}

The CS(2-1) and CS(3-2) lines trace even denser molecular gas at
10$^5$cm$^{-3}$. The problem is that CS line emission is quite
weak. Detections of CS in extra-galactic sources are sparse. In
the sample of normal and interacting galaxies and ULIRGs Solomon
et al.(1992) detected CS line emission only in Arp~220. For Cen~A
we find a total CS(3-2) line luminosity of $L_{\rm CS(3-2)} \sim
5.4 \times 10^5 {\rm K~km~s^{-1} pc^2}$. The ratio of the CS to CO
line luminosity is then $L_{\rm CS(3-2)}/L_{\rm CO}\sim 1/170$,
very similar to the value of 1/250 given by Solomon et al. (1990)
for the Milky Way. For the Galactic Center the same authors derive
a corresponding ratio of about 1/20 to 1/30. This indicates that
the Galactic Center is approximately 8 to 10 times more luminous
in the CS(3-2) line that the molecular gas in the disk. At the
center of Cen~A the $L_{\rm CS(3-2)}$ is 2 times higher compared
to the off-center dust lane measurement.

The CS(3-2) line luminosity results in a total mass estimate of
gas at densities of 10$^5$cm$^{-3}$ of about $M_{\rm
CS(3-2)}(H_2)$ = (2-8)$\times$10$^7$M$_{\odot}$. This number is in
very good agreement with the corresponding value derived from the
HCN(1-0) line emission.

\section{Conclusion}

While the line luminosities of the HCN(1-0) and CS(3-2) as well
as the FIR luminosity of the molecular gas in the dust lane of
Cen~A are quite comparable to each other there are definite
differences in the overall fraction of dense molecular gas and the
efficiency with which stars are formed from it. This star
formation activity is also the source of the FIR luminosity of
Cen~A. About 40\% or even more of the total molecular line
luminosity in Cen~A originates in dense gas. This suggests that
star formation as well as the bulk of the dense molecular gas is
mostly concentrated in GMC complexes rather than in a more diffuse
molecular gas component. This is already indicated by the
ring-like distribution of HII region found by Graham (1979) as
well as the MEM deconvolved $^{12}$CO(2-1) line emission mass by
Rydbeck et al. (1993).

We also note that with respect to other positions in the dust lane
the I(HCN)/I(CO) ratio is larger at separations of about
100$^{\prime \prime}$ from the nucleus rather than at separations
of about 60$^{\prime \prime}$. The larger distance is close to the
inner edge of the ring of HII regions and corresponds well with
the position of the folds in the warped molecular gas disk of
Centaurus~A (Quillen et al. 1992, 1993; Sparke 1996) and an increased
intensity in the 15$\mu$m continuum dust emission (Block \&
Sauvage 2000; Mirabel et al. 1999). Therefore a higher
I(HCN)/I(CO) ratio may be due to a combination of enhanced star
formation efficiency at these positions and an increase in column
density due to the folds. Due to the low intrinsic velocity
dispersion of the thin molecular disk (Quillen et al. 1992) and
due to the fact that the $^{12}$CO line emission is originating in
optically thick molecular gas (Wild et al. 1997) this
may lead to shadowing of molecular clouds along the line of sight
toward the folds. This effect will be stronger for $^{12}$CO than
for the small, dense cloud cores seen in the less abundant HCN
line emission. This effect may therefore lead to an intensity
decrease in the $^{12}$CO line and an increase in the HCN(1-0)
line, resulting in the observed variation of the I(HCN)/I(CO) line
ratio.

Future interferometric measurements will allow us to study the
distribution of molecular gas in the dust lane of Centaurus~A in
much greater detail. Line ratios, luminosities and star formation
can then be investigated for individual GMC complexes.


\begin{acknowledgements}
We are grateful to the SEST team and the ESO staff on La Silla and
in Garching for their support and hospitality. We thank Lars-\AA
ke Nyman for taking an additional central CO spectrum.
\end{acknowledgements}


\begin{thebibliography}{}

\bibitem[1988]{bell}Bell M.B., Seaquist E.R., 1988, ApJ 329, L17
\bibitem[2000]{block}Block D.L., Sauvage M., 2000, A\&A 353, 72
\bibitem[1983]{ebneter}Ebneter K., Balick B, 1983, PASP 95, 675
\bibitem[1990a]{eckart}Eckart A.,Cameron M., Genzel R., Jackson J., Rydbeck G., 1990a,
ApJ 365, 522
\bibitem[1990b]{eckart}Eckart A., Cameron M., Rothermel H. et al., 1990b, ApJ 363, 451
\bibitem[1976]{gardner}Gardner F.F.,  Whiteoak J.B., 1976,
Proc. Astron. Soc. Aust. 3, 63
\bibitem[1979]{graham}Graham J.A., 1979, ApJ 232, 60
\bibitem[1993]{helfer}Helfer T.T., Blitz L., 1993, ApJ
419, 86
\bibitem[1995]{helfer}Helfer T.T., Blitz L., 1995,
ApJ 450, 90
\bibitem[1993]{hui}Hui X., Ford H.C., Ciardullo R., Jacoby G.H., 1993, ApJ 414, 463
\bibitem[1983]{van der hulst}van der Hulst J.M., Golish W.F., Hashick A.D.,
       1983, ApJ 264, L37
\bibitem[1992]{israel}Israel F.P., 1992, A\&A 265, 487
\bibitem[1998]{israel}Israel F.P., 1998, A\&AR 8, 237
\bibitem[1990]{israel}Israel F.P., van Dishoeck E.F., Baas F. et al., 1990, A\&A 227, 342
\bibitem[1991]{israel}Israel F.P., van Dishoeck E.F., Baas F.,
de Graauw Th., Phillips T.G., 1991, A\&A 245, L13
\bibitem[1993]{jackson}Jackson J.M., Paglione T.A.D., Ishizuki S.,
Nguyen-Q-Rieu, 1993, ApJ 418, L13
\bibitem[1988]{joy}Joy M., Lester D.F., Harvey P.M., Ellis H.B., 1988, ApJ 326, 662
\bibitem[1974]{kellermann}Kellermann K.I., 1974, ApJ 194, L135
\bibitem[1997]{kellermann}Kellermann K.I., Zensus A., Cohen M.H., 1997, ApJ 475, L93
\bibitem[1996]{kohno}Kohno K., Kawabe R., Tosaki T., Okumura S.K., 1996, ApJ 461, L29
\bibitem[1981]{kutner}Kutner M.L., Ulich B.L., 1981, ApJ 250, 341
\bibitem[1975]{kwan}Kwan J., Scoville N., 1975, ApJ 195, L85
\bibitem[1990]{lee}Lee Y., Snell R.L., Dickman R.L., 1990, ApJ 355, 536
\bibitem[1980]{linke}Linke R.A., Goldsmith P.F., 1980, ApJ 235, 437
\bibitem[1988]{marston}Marston A.P, Dickens R.J., 1988, A\&A 193, 27
\bibitem[1999]{mirabel}Mirabel I.F., Laurent O., Sanders D.B., 1999, A\&A 341, 667
\bibitem[1992]{nguyen}Nguyen-Q-Rieu, Jackson J.M., Henkel C., Truong B., Mauersberger R., 1992, ApJ 399, 521
\bibitem[1987]{phillips}Phillips T.G., Ellison B.N., Keene J.B. et
al., 1987, ApJ (Letters) 322, L73
\bibitem[1992]{quillen}Quillen A.C., de Zeeuw P.T., Phinne, E.S., Phillips T.G., 1992,
ApJ 391, 121
\bibitem[1993]{quillen}Quillen A.C., Graham J.R., Frogel J.A., 1993, ApJ 412, 550
\bibitem[1993]{rydbeck}Rydbeck G., Wiklind T., Cameron M. et al., 1993, A\&A 270, L13
\bibitem[1993]{sage}Sage L., Galleta G., 1993, ApJ 419, 544
\bibitem[1986]{seaquist}Seaquist E.R., Bell M.B., 1986, ApJ 303, L67
\bibitem[1990]{seaquist}Seaquist E.R., Bell M.B., 1990, ApJ 364, 94
\bibitem[1975]{shaffer}Shaffer D.B., Schilizzi R.T., 1975, ApJ 80, 753
\bibitem[1996]{sparke}Sparke L.S., 1996, ApJ 473, 810
\bibitem[1996]{soria}Soria R., et al., 1996, ApJ 465, 79
\bibitem[1990]{solomon}Solomon P.M., Radford S.J.E., Downes D., 1990, ApJ 348, L53
\bibitem[1992]{solomon}Solomon P.M., Downes D., Radford S.J.E., 1992, ApJ 387, L55
\bibitem[1994]{sternberg}Sternberg A., Genzel R., Tacconi L., 1994, ApJ 436, L131
\bibitem[1994]{tacconi}Tacconi L.J., Genzel R., Blietz M. et al., 1994, ApJ 426, L77
\bibitem[1979]{de vaucouleurs}de Vaucouleurs G., 1979, AJ 84, 1270
\bibitem[1971]{whiteoak}Whiteoak J.B., Gardner F.F., 1971, ApJ 8, L57
\bibitem[1997]{wiklind}Wiklind T., Combes F., 1997, A\&A 324, 51
\bibitem[1997]{wild}Wild W., Eckart A., Wiklind T., 1997, A\&A 322, 419

\end{thebibliography}
\end{document}